\documentclass{article}
\usepackage{waspaa21,amsmath,graphicx,url,times}
\usepackage{tabularx, makecell,array,booktabs, multirow, amssymb, graphicx, cite}
\newcolumntype{P}[1]{>{\centering\arraybackslash}p{#1}}

\usepackage{color}

\newcommand{\fref}[1]{Figure~\ref{#1}}
\newcommand{\tref}[1]{Table~\ref{#1}}
\newcommand{\sref}[1]{Section~\ref{#1}}

\title{Who calls the shots? Rethinking few-shot learning for audio}

\name{Yu Wang,$^{1}$\sthanks{This work is partially supported by NSF award 1544753.}
      Nicholas J. Bryan$^{2}$,
      Justin Salamon$^{2}$,
      Mark Cartwright,$^{3}$
      Juan Pablo Bello$^1$}
\address{$^1$ Music and Audio Research Laboratory, New York University, New York, NY, USA,\\
         $^2$ Adobe Research, San Francisco, CA, USA\\
$^3$ New Jersey Institute of Technology, Newark, NJ, USA\\ 
}

\begin{document}

\ninept
\maketitle

\begin{sloppy}

\begin{abstract}

Few-shot learning aims to train models that can recognize novel classes given just a handful of labeled examples, known as the support set.
While the field has seen notable advances in recent years, they have often focused on multi-class image classification. Audio, in contrast, is often multi-label due to overlapping sounds, resulting in unique properties such as polyphony and signal-to-noise ratios (SNR). This leads to unanswered questions concerning the impact such audio properties may have on few-shot learning system design, performance, and human-computer interaction, as it is typically up to the user to collect and provide inference-time support set examples.
We address these questions through a series of experiments designed to elucidate the answers to these questions. We introduce two novel datasets, FSD-MIX-CLIPS and FSD-MIX-SED, whose programmatic generation allows us to explore these questions systematically. Our experiments lead to audio-specific insights on few-shot learning, some of which are at odds with recent findings in the image domain: there is no best one-size-fits-all model, method, and support set selection criterion. Rather, it depends on the expected application scenario. Our code and data are available at \url{https://github.com/wangyu/rethink-audio-fsl}.

\end{abstract}


\section{Introduction}
\label{sec:intro}

The field of few-shot learning, i.e., training a model such that it can recognize previously unseen classes given a small set of examples, has seen significant progress in recent years, especially in the image domain~\cite{Koch2015, finn2017, ravi2017, snell2017, sung2018, chen2019}. Advances have also been made in the audio domain~\cite{pons2019, Zhang2019, chou2019, wang_icassp_2020, wang_ismir_2020, shi2020, shimadaKI2020}, including for few-shot continual learning where a model can recognize a predefined set of base classes \emph{and} learn to recognize new classes given few examples~\cite{wang_icassp_2021}, but many challenges still remain for this domain.

Audio data have unique characteristics that set them apart from image data. Most importantly, sound events may overlap, leading to attributes such as polyphony (i.e., how many sounds are overlapped), per-sound-event signal-to-noise (foreground to background) ratios, and weakly labeled data. While the computer vision literature has focused mostly on strongly labeled, single-label images (i.e., multi-class problems)~\cite{Koch2015, finn2017, ravi2017, snell2017, chen2019}, many audio recognition problems like sound event detection (SED), instrument recognition, and speaker recognition are often multi-label.

The multi-label nature of many audio recognition problems complicates the inference-time support set example selection process for users of few-shot learning systems. It is no longer merely a question of how many examples to provide (i.e., the size of the \emph{support set}). Now, users must determine how to compose the best support set in terms of each sample's polyphony and SNR. For example, is it okay to provide examples with other overlapping sounds?  Should the support set only contain a given target sound?  How does the background noise level of the support set affect performance?

\begin{figure}[t]
\centering
\includegraphics[clip, trim=10.5cm 8.5cm 11cm 3.5cm, width=\columnwidth]{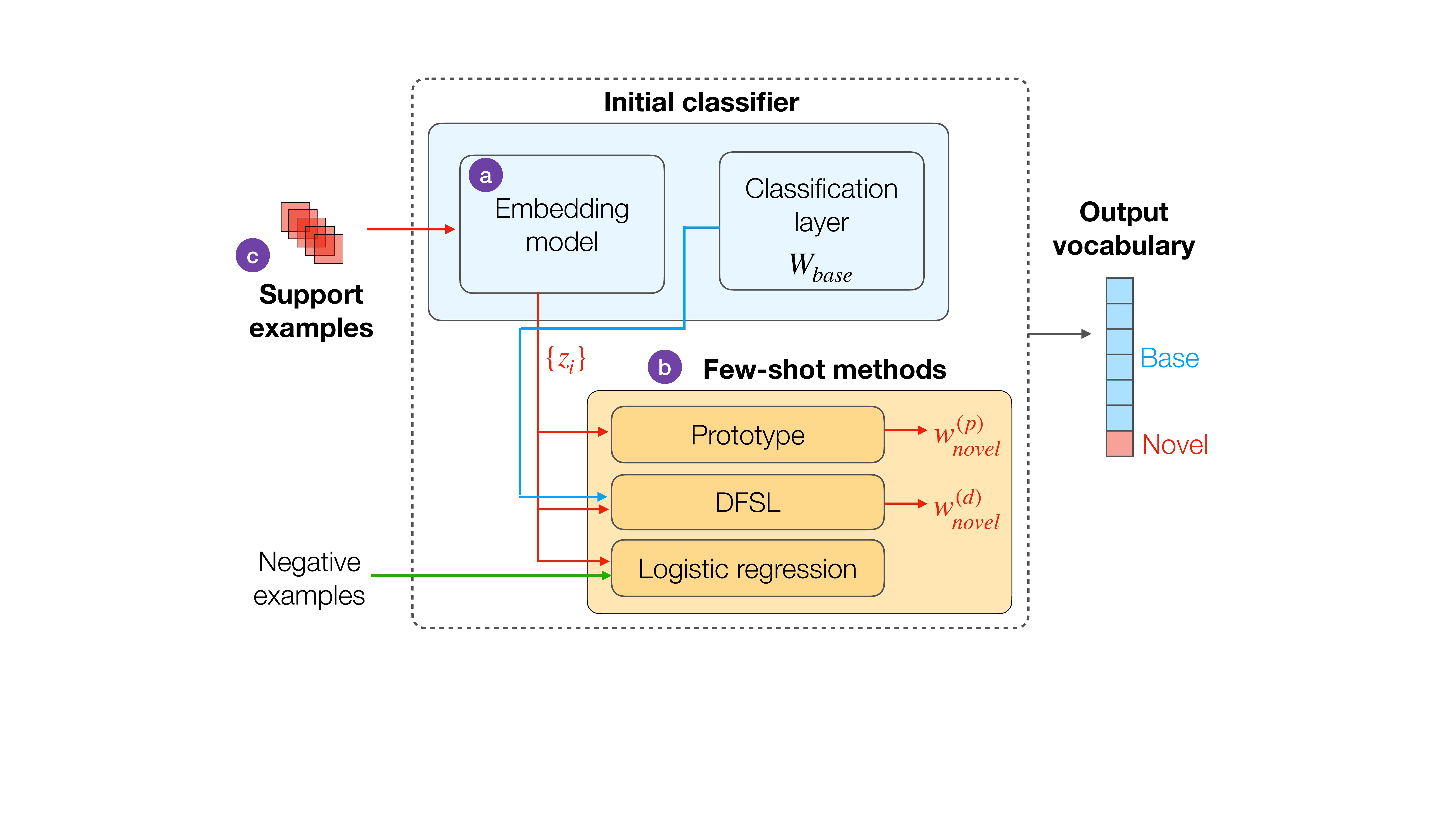}
\caption{Few-shot continual learning framework for multi-label audio classification. We analyze (a) embedding model architectures, (b) few-shot methods, and (c) support set selection.}
\label{fig:framework}
\end{figure}

In this paper, we elucidate the relationship between few-shot model design (architecture), optimization strategy (few-shot method), and inference-time support set example selection within the framework of few-shot continual learning for multi-label audio classification, as shown in \fref{fig:framework}. 
Our analysis leads us to insights that are unique to the audio domain, sometimes in contradiction with recent findings in computer vision~\cite{tian2020rethinking}. Our contributions can be summarized as follows:
\begin{itemize}
    \item We introduce novel datasets, FSD-MIX-CLIPS and FSD-MIX-SED, allowing us to study the impact of audio-specific qualities (polyphony, SNR) on few-shot learning in a controlled way. 
    \item We show that there are design choices between few-shot models, methods, and inference-time support set selection and that there is currently no one-size-fits-all solution. Rather, it depends on the anticipated application scenario that includes desired user labeling effort, runtime computation and storage resources, and prior knowledge of the test data.
    \item We provide audio-specific insights for few-shot learning including guidance on the choice of support set in terms of size, polyphony, and sound event SNR that can be used for both system design and motivate new few-shot learning methods.
\end{itemize}
\section{The FSD-MIX-CLIPS and FSD-MIX-SED Datasets}
\label{sec:dataset}

We begin by surveying past audio datasets, including ESC-50~\cite{Piczak2015}, AudioSet~\cite{gemmeke2017audio} and FSD50K~\cite{fonseca2020fsd50k}. ESC-50 contains a small set of environmental recordings of 50 classes.
It is high-quality, but also too simple and unrealistic for our needs as it contains single-source audio and minimal background noise. In contrast, AudioSet is a large-scale dataset of over 2M human-labeled 10s sound clips drawn from YouTube videos with 527 sound classes. AudioSet, however, is not fully open for audio download
and the annotation quality varies notably between 
classes\footnote{\label{audioset-quality}\url{https://research.google.com/audioset/dataset/index.html}}. Such issues motivated the development of FSD50K~\cite{fonseca2020fsd50k}, which is a fully open dataset that contains over 51k audio clips manually labeled using 200 classes drawn from the AudioSet Ontology and content gathered from Freesound~\cite{font2013freesound}.  
Unfortunately, FSD50K does not provide strong labels of events across time and polyphony, and does not contain SNR annotations -- both central to our study. Thus, we build upon this work to construct a programmatically-mixed dataset, FSD-MIX-CLIPS, in which we control the polyphony and SNR characteristics. \tref{tab:fsd-sed} shows the number of classes and mixed clips in each split in FSD-MIX-CLIPS, as well as the number of single-labeled clips from FSD50K used as source material for mixing. In a few-shot continual learning framework, we need disjoint sets of \textit{base} and \textit{novel} classes where novel class data are only used at inference time. 

To generate clips with multiple sound events, we first filter FSD50K to get a set of single-labeled clips as the foreground sound material. We choose clips shorter than 4s that have a single validated label with the \textit{Present and Predominant} annotation type~\cite{salamon2014dataset, fonseca2020fsd50k}. We further trim the silence at the edges of each clip using pysox~\cite{bittner2016pysox} with a threshold of 0.1\% and a minimum duration of 0.01s. The resulting subset contains clips each with a single and strong label. The 200 sound classes in FSD50K are hierarchically organized. We focus on the leaf nodes and rule out classes with less than 20 single-labeled clips. This gives us 89 sound classes and a total of around 10k single-labeled clips. Next, we partition the 89 classes into three splits: \textit{Base}, \textit{Novel-val}, and \textit{Novel-test} with 59, 15, and 15 classes, respectively. For base classes, we follow the original data split in FSD50K, where clips in the development set are used for training and validation with a ratio of 5:1, and clips in the evaluation set are used for testing. Novel classes are only used at inference time during validation or testing. Thus, for each novel class, we combine the clips from the development and evaluation sets of FSD50K.

We generate 10-second strongly-labeled soundscapes with controlled SNR and polyphony with Scaper~\cite{salamon2017scaper}. We use the single-labeled FSD50K clips as foreground sounds and Brownian noise as the background. We define the number of classes $c$ in each soundscape by drawing from 1 to 5 with probability proportional to $1/c$. Then we sample $c$ classes from one of the class splits without replacement. 
From each of the $c$ classes, we sample one clip as a foreground sound event. 
Each sound event is randomly pitch-shifted within +/- 2 semitones and time-stretched by a ratio in [0.8, 1.2]. 
All transformed clips are then randomly placed in a soundscape and mixed with an SNR randomly sampled from [-5, 20] dB. 
We refer to this (intermediate) dataset of 10s soundscapes as FSD-MIX-SED.



Finally, we extract 615k 1s clips centered on each sound event in the soundscapes in FSD-MIX-SED to produce FSD-MIX-CLIPS. To label a clip, we consider all sound events within the 1s window. If an event overlaps with the window for more than 0.5s or half of the event duration, we add the corresponding class into the clip label. We then consider the number of classes within a clip as the level of polyphony with the assumption that it is rare to have short non-overlapping events within a 1s window. 

One of our goals is to study the impact of support set polyphony and SNR, and it is cleaner to do so under a multi-label audio classification setting (i.e., tagging), as opposed to SED where post-processing plays an important factor. For this reason, the remainder of this work will focus on FSD-MIX-CLIPS. By making both datasets available, we hope to encourage researchers to build on this work, e.g., by translating this study to a full SED setting.


\begin{table}[t]
\centering
\footnotesize
\begin{tabular}{c|P{1.2em}P{1.2em}P{1em}P{4.2em}P{4.3em}}
\toprule
\textbf{Class split}& \multicolumn{3}{c}{\textbf{Base}} & \textbf{Novel-val} & \textbf{Novel-test} \\
\midrule
\# Classes & & 59  &  & 15 & 15 \\
\midrule
\midrule
\textbf{Data split} & \textbf{Train} & \textbf{Val.} & \textbf{Test} & \textbf{Val.} & \textbf{Test} 
\\
\midrule
FSD-MIX-SED (10s) & 200k & 30k  & 30k  & 8k & 8k \\
FSD-MIX-CLIPS (1s) & 450k & 65k  & 65k  & 17k & 17k \\
FSD50K clips used & $\sim$5k & $\sim$1k  & $\sim$2k  & $\sim$1k & $\sim$1k \\
\bottomrule
\end{tabular}
\caption{Numbers of classes and mixed clips/soundscapes in FSD-MIX-CLIPS (1s) and FSD-MIX-SED (10s) per split and the corresponding number of single-labeled clips used from FSD50K.}
\label{tab:fsd-sed}
\end{table}

\section{Experimental Design}
\label{sec:exp}
We perform a series of experiments with FSD-MIX-CLIPS to understand design choices between few-shot model architectures, learning methods, and support set selection under the few-shot continual learning framework for multi-label audio classification. Moreover, to clearly elucidate support set selection issues, we specifically analyze  
different support set compositions, including the number of support examples, polyphony, and SNR.

\subsection{Embedding model architectures}
\label{sec:}
We first investigate two common convolutional neural network (CNN) model architectures, \textit{Conv4} and \textit{PANN}, to understand how embedding model architecture and capacity affect overall few-shot learning performance. 
\textit{Conv4} consists of four convolution blocks, each of which has a convolutional layer with a $3 \times 3$ kernel, a batch normalization layer, a ReLU activation layer, and a $2 \times 2$ max-pooling layer. The first two convolutional layers have 64 feature channels and the latter two have 128 feature channels. We flatten the last feature map to get output embeddings with dimension 3072 and $\approx440$k trainable parameters. On the other hand, \textit{PANN} is a powerful 14-layer CNN~\cite{kong2020panns}, which achieved state-of-the-art audio tagging results. The output embedding has 2048 dimensions, and the number of trainable parameters is $\approx$80M.  

Besides training the embedding model from scratch, we also investigate using a pre-trained OpenL3 embedding model~\cite{arandjelovic2017look,  cramer2019look}. Here, we seek to understand if pre-training through self-supervised learning of audio-visual correspondence in YouTube videos provides a more generalizable representation.  We extract OpenL3 embeddings with dimension 512 from 1s audio inputs, and train an additional fully-connected layer to transform the embedding dimension to 2048 with $\approx$1M trainable parameters. We refer to this model as \textit{pre-OpenL3+FC}. To disentangle the effect of pre-training and model capacity, we also train an embedding model with the \textit{OpenL3+FC} architecture from scratch without pre-training. 


\subsection{Few-shot methods}

\label{sec:exp-method}
We investigate three few-shot methods, \textit{Prototype}, \textit{DFSL}, and \textit{LR}, to understand how different learning paradigms affect performance and their relationship to the support set. To predict a novel class based on few data, both the \textit{Prototype} and \textit{DFSL} approaches aim to compute a new classification weight vector to add to the existing base weight matrix as shown in \fref{fig:framework}(b), while \textit{LR} aims to learn a new binary logistic regression model directly.  

The \textit{Prototype} approach~\cite{gidaris2018dynamic, snell2017} treats the averaged embedding $z$ of the support examples for a novel class as a prototype, and uses it as a novel classification weight vector: $w^{(p)}_{novel} = z_{avg} = \frac{1}{n}\sum_{i=1}^{n}z_{i}$, where $n$ is the number of support examples. This approach is a simple way to extend the initial classifier to novel classes that relies solely on the embedding model. Dynamic Few-Shot Learning (\textit{DFSL})~\cite{gidaris2018dynamic, wang_icassp_2020} goes a step further to train a few-shot classification weight generator to generate novel classification weight vectors. The weight generator not only takes in the prototype of a novel class $z_{avg}$ but also exploits past knowledge of the initial classifier by incorporating a cosine-similarity based attention mechanism over the base classification weight matrix $W_{base}$ to get $w^{(d)}_{novel} = \phi_{avg} \odot z_{avg} + \phi_{att} \odot w'_{att}$.
Here, $w'_{att}$ can be viewed as a learned weighted sum of base class weight vectors in $W_{base}$. $\phi_{avg}$ and $\phi_{att}$ are learnable weights, and $\odot$ is the Hadamard product. The few-shot weight generator is trained via episodic training~\cite{vinyals2016}, where each training iteration mimics the testing scenario. First, 5 ``pseudo" novel classes are sampled from the base classes. We treat each pseudo novel class as if it is an actual novel class at inference time, and sample $n$ examples from it. Then we generate new weight vectors for these pseudo-novel classes based on support examples and $W_{base}$.
We replace the classification weight vectors of pseudo-novel classes temporally and update the few-shot weight generator parameters to minimize the classification loss on a batch of data with both base and pseudo-novel classes.



Lastly, recent few-shot image classification work has shown that a simple logistic regression or \emph{LR} model trained on top of a pre-trained embedding outperforms many meta-learning based algorithms \cite{tian2020rethinking}. It suggests that good learned embeddings are capable of fine-tuning on few data. We evaluate this approach to see if it is as effective for multi-label audio classification under a few-shot continual learning setup. At inference time, for each novel class, we train a binary LR model on the support example embeddings. Note that because the support examples are labeled examples of a novel class,  we also need negative examples to train a binary classifier. Since the training data does not include any of the novel classes, we randomly sample $x$ examples from the training set as negative examples, where $x$ is a hyperparameter tuned on the validation set.

\subsection{Support set selection}
\label{sec:}
Support examples play an important role in the few-shot continual learning paradigm, as they are the only supervision given for each novel class. 
So, we vary the number of support examples $n$ 
from 5, 10, 20, to 30, representing standard few-shot and low data scenarios. Then, given $n$, we experiment with audio-specific characteristics, polyphony and SNR, of the support examples. First, we experiment with \textit{monophonic} and \textit{polyphonic} support sets. \textit{Monophonic} means we only include examples containing a single sound event from the target novel class. \textit{Polyphonic} 
means the examples may contain, in addition to the target class, other overlapping sounds.
Next, we experiment with two versions, \textit{low-SNR} and \textit{high-SNR}, of monophonic support sets. We define the SNR values within [-5, 0, 5] (dB) as \textit{low-SNR}, and [10, 15, 20] (dB) as \textit{high-SNR}.   

Note that for few-shot methods that involve model training with support examples at either train time or inference time (\textit{DFSL} and \textit{LR}), the support set characteristics are matched between training and testing in our experiments.  

\subsection{Training and evaluation setups}
Given our FSD-MIX-CLIPS dataset, embedding model architectures, few-shot learning methods, and a set of distinct support set characteristics, we start by training the initial classifier on the \textit{Base} training set. We downsample each audio clip to 16 kHz and compute a 64-bin log-scaled Mel-spectrogram as the input to the model using librosa\cite{mcfee2019} with a 25 ms window length, 10 ms hop size, and fast Fourier transform size of 64 ms. We implement the classifier in PyTorch~\cite{paszke2019} and use the Adam optimizer~\cite{kingma2014} with a 0.001 learning rate with early stopping. Then, if \textit{DFSL} is used as the few-shot method, we further train the few-shot classification weight generator in an additional training stage, also on the \textit{Base} training set. 

At evaluation, for each class in the \textit{Novel-test} split, we sample a set of support examples. By combining the remainder of the novel class data with the test data for base classes, we get a test set which is a mixture of base and novel class samples. If \textit{LR} is used as the few-shot method, we train one binary logistic regression model for each novel class using the support examples. We use scikit-learn~\cite{pedregosa2011scikit} implementation for the logistic regression model with balanced class weights and a maximum of 1000 iterations. Finally, we combine the initial base classifier and the few-shot method to predict the test set labels in a joint label space. We run 100 iterations of this evaluation procedure to account for sampling randomness, and compute averaged per-class F-measure with a fixed threshold of 0.5. We aggregate the per-class F-measure and compute the mean over base and novel classes separately.
In addition to computing metrics based on class labels, the automatic annotation produced by Scaper for FSD-MIX-CLIPS allows us to break down model performance by test-set polyphony, ranging from 1~to~4, and by SNR, ranging from -5 to 20 dB.

\section{Results}
\label{sec:result}

\subsection{Embedding model architectures}
\label{sec:model}
In \fref{fig:model}, we show the performance on base and novel classes with different embedding models. All other variables are kept fixed in this set of experiments, with \textit{DFSL} as the few-shot method, monophonic support examples with mixed SNR, and $n=5$. 

First, we see that the base class performance increases with increasing model capacity. This matches our intuition that standard multi-label classification benefits from a more powerful model. Second, this trend does not hold for novel classes. \textit{PANN} achieves the highest base class performance but low novel class performance, suggesting a trade-off between overfitting base classes and generalizing to novel classes. Third, \textit{pre-OpenL3+FC} achieves the best novel class performance and the most balanced performance between base and novel classes. This can potentially be attributed to pre-training as \textit{OpenL3+FC}, which is trained from scratch, shows lower novel class performance. We conjecture pre-training helps prevent overfitting to base classes and improves generalization to novel classes--this model has ``seen'' significantly more data during pre-training 
via self-supervision on unlabeled data.
We fix the embedding model to \textit{pre-OpenL3+FC} in the rest of the work.

\begin{figure}[t]
\centering
\includegraphics[trim=0 0 0 0, width=\columnwidth]{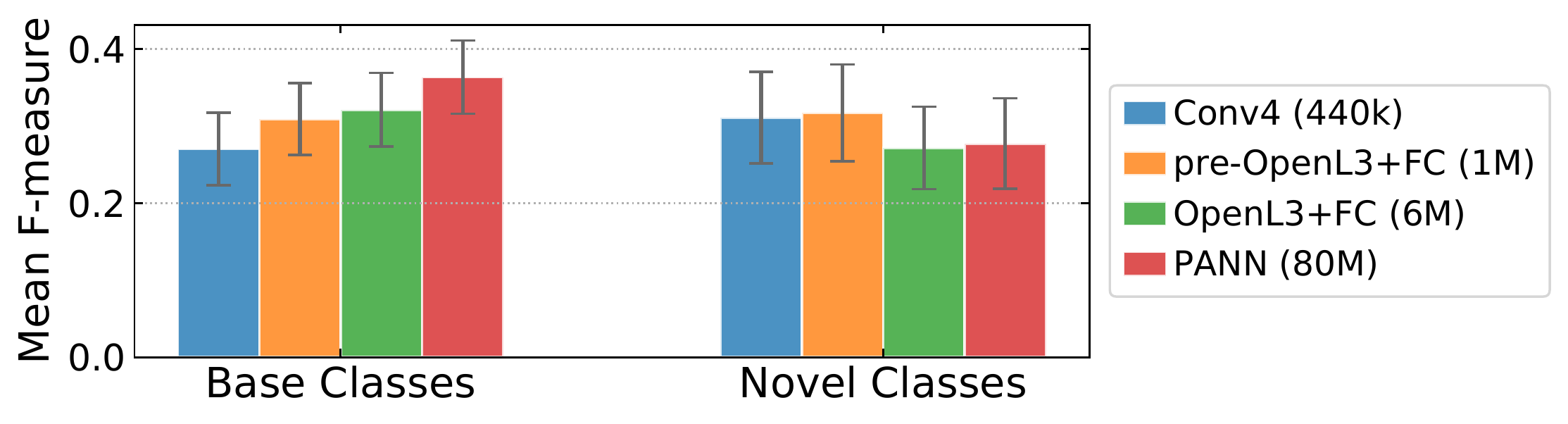}
\vspace{-.5cm}
\caption{Mean F-measures averaged over 59 base classes and 15 novel classes. The number of learnable parameters are shown after model names. Error bars represent 95\% confidence intervals.}
\label{fig:model}
\end{figure}

\subsection{Few-shot methods and number of support examples}
\label{sec:metho}
Next, we look at the performance of different few-shot methods on predicting novel classes with $n$ varied from 5 to 30, representing standard few-shot to low data scenarios. Here, the support sets are monophonic with mixed SNR. The results, shown in \fref{fig:method}, provide several insights.
First, we find that \textit{DFSL} performs the best out of all methods when $n=5$, a typical few-shot learning scenario. Among the three methods we experimented with, \textit{DFSL} was specifically designed to solve few-shot learning problems as it leverages an additional episodic training stage to train the few-shot weight generator. Interestingly, this finding is in contrast to the conclusions reached for single-label, multi-class, few-shot image classification~\cite{tian2020rethinking}. Second, as $n$ increases to 10 or more, \textit{LR} starts to outperform \textit{DFSL}. This indicates that a simple transfer learning approach becomes more effective when the number of support examples increases. The downside of \textit{LR}, however, is that it requires additional optimization and data storage at inference time, since we need to train a binary logistic regression model for each novel class. Training such binary models, as mentioned in \sref{sec:exp-method}, requires additional negative examples randomly sampled from the training data. The optimal number of negative examples at different $n$, based on validation performance, ranges from 100 to 5000. We need to store the embeddings of these negative examples to train each binary model. On the other hand, while \textit{DFSL} requires an additional training stage, it only requires a forward pass at inference time to generate novel classification weight vectors to predict novel classes.

We can formulate these findings into the following design choices. When there is no constraint on annotation budget or computation and storage resources, \textit{LR} as the few-shot method with large $n$ is the best combination. However, if it is critical to minimize user labeling or runtime resources, \textit{DFSL} with small $n$ is preferable.   

\begin{figure}[t]
\centering
\includegraphics[trim=0 0 0 0, width=\columnwidth]{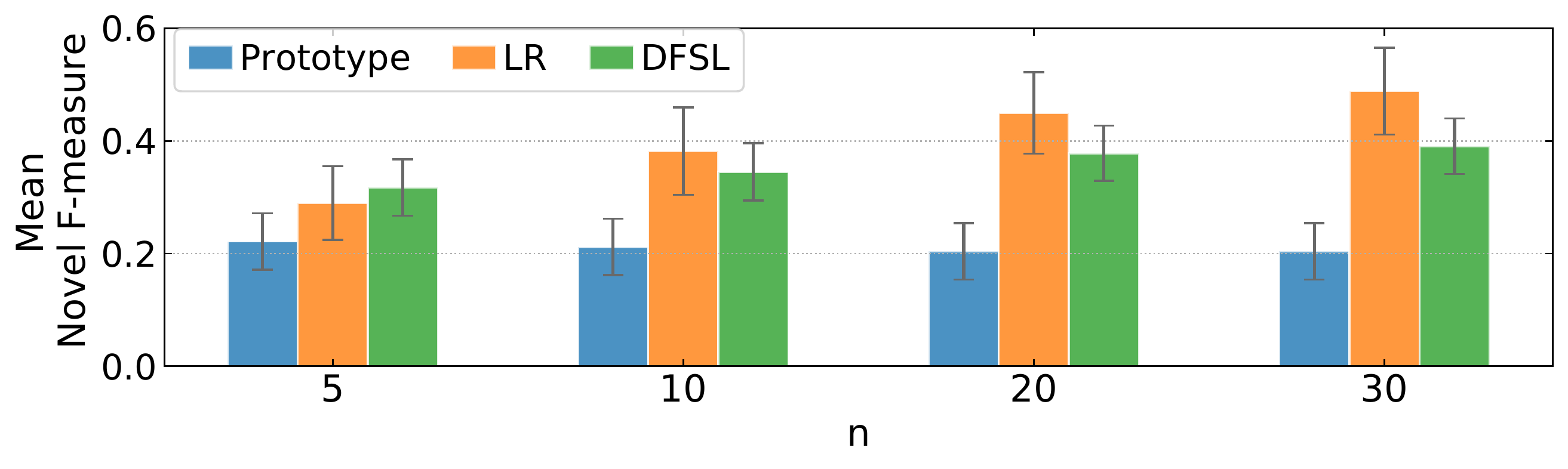}
\vspace{-.7cm}
\caption{Mean F-measures averaged over 15 novel classes. Error bars represent 95\% confidence intervals.}
\label{fig:method}
\end{figure}

\subsection{Polyphony and SNR of support and test samples}
\label{sec:poly}
Lastly, we show how the polyphony and SNR of the support set affect model performance when predicting novel classes in test samples. Here, we focus on the $n=5$ few-shot scenario for both \textit{LR} and \textit{DFSL}, but we found similar trends for $n=30$.

First, the results, displayed in \fref{fig:poly-snr}, show that matching the support set characteristics to those of the test samples, both in terms of polyphony and SNR, gives the best results.
For both few-shot methods, the best performance on monophonic and highly polyphonic test samples is achieved by monophonic and polyphonic support sets, respectively. Similarly, the best performance on test samples with low/high SNR is achieved by support sets with low/high SNR, respectively. Second, \fref{fig:poly-snr} (left) shows that polyphonic support sets lead to more consistent performance across test set polyphonies, while the highest overall performance comes from monophonic support sets on monophonic test samples. \fref{fig:poly-snr} (right) shows that support sets with low SNR lead to more consistent test performance across SNR, while support sets with high SNR achieve the highest recall on high-SNR test samples. Lastly, we see that \textit{LR} exhibits a dramatic performance drop when there is a mismatch between the polyphony of the support and test samples. A possible explanation is that each binary LR model is trained directly on the support examples, and therefore, the model performance is more dependent on support example characteristics. Whereas for \textit{DFSL}, the base weight matrix is included as additional information.

We can distill these observations into two reproducible insights for our multi-label few-shot continual learning audio classification task. First, if we know the test sample characteristics (e.g., mostly monophonic with high SNR), matching those in the support set will lead to better performance. Conversely, if we do not have prior knowledge about the test data, opting for support examples with more complex acoustic characteristics (i.e., mixed polyphony and lower SNR) would be a safer choice which leads to more consistent performance. This holds for both the polyphony and SNR characteristics, using either the \textit{DFSL} or \textit{LR} learning methods. 

\begin{figure}[t]
\centering
\includegraphics[trim=0 0 0 0, width=\columnwidth]{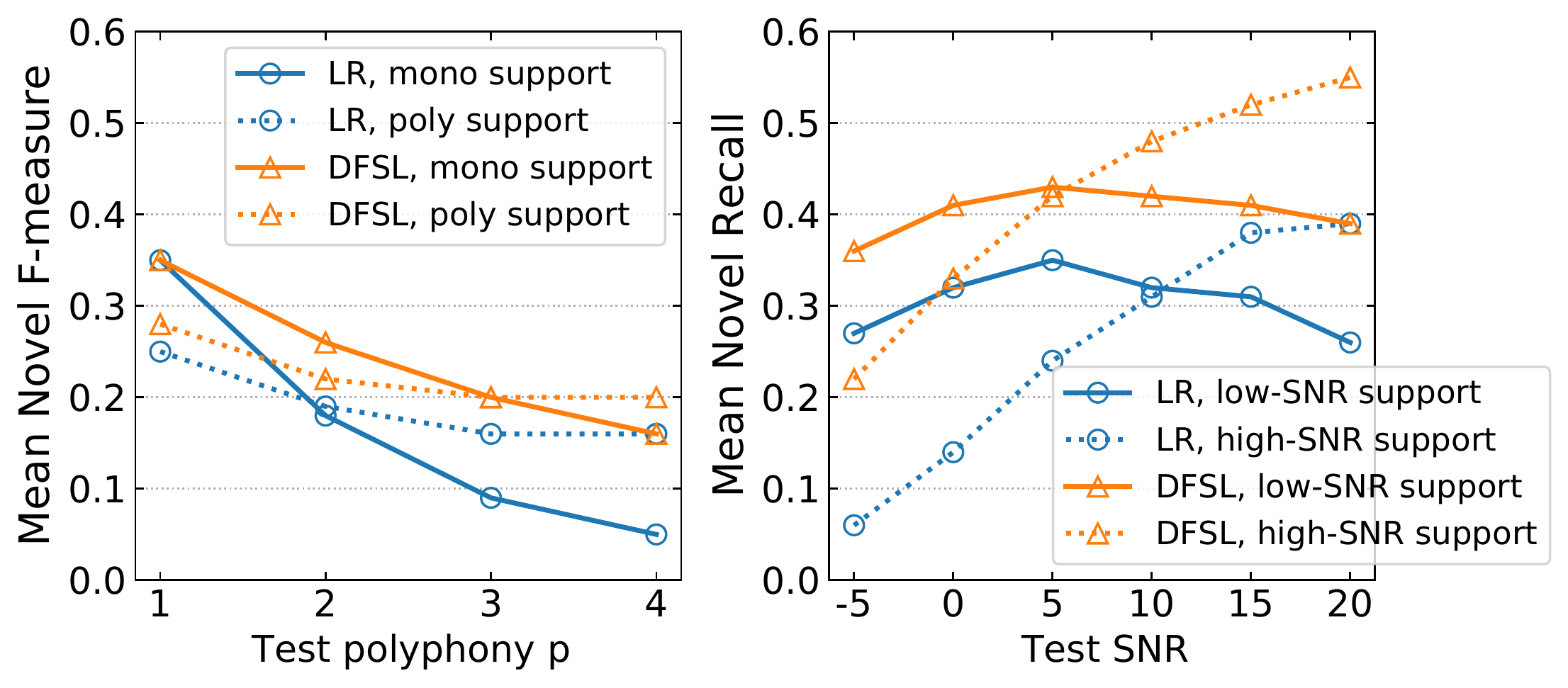}\vspace{-2mm}
\caption{Performance breakdown by (left) polyphony $p$ and (right) SNR of test samples. Note that we compute recall in the right plot since we can only define SNR when the sound is present.}
\label{fig:poly-snr}
\vspace{-1mm}
\end{figure}


\section{Conclusions}
\label{sec:}
In this work, we study few-shot learning for audio. We create a programmatically-mixed dataset, FSD-MIX-CLIPS, to help us to control important common audio data characteristics including polyphony and SNR. Then, we perform a series of experiments to elucidate how model architecture, learning strategy, and support set selection affect performance of few-shot continual learning on multi-label audio classification. Generalizing to real-audio datasets is a potential future work when datasets with comparable size and detailed annotation become available. Our results lead us to useful audio-specific insights, some of which are at odds with recent findings in the image domain: there is no current one-size-fits-all best performing approach, but rather, design choices should depend on the expected application scenario.


\bibliographystyle{IEEEtran}
\bibliography{main}

\end{sloppy}
\end{document}